\documentclass[twocolumn,showpacs,preprintnumbers,showkeys,superscriptaddress]{revtex4}  %
\usepackage{amssymb}
\usepackage{graphicx}
\usepackage{dcolumn}
\usepackage{bm}
\usepackage{appendix}
\usepackage{epstopdf}

\def\eqref#1{Eq.~(\ref{eq:#1})}

\begin{document}

\title{Generalized Seniority on Deformed Single-Particle Basis}
\author{L. Y. Jia}  \email{liyuan.jia@usst.edu.cn}
\affiliation{Department of Physics, University of Shanghai for
Science and Technology, Shanghai 200093, P. R. China}

\date{\today}


\begin{abstract}

Recently we proposed \cite{Jia_2015} a fast computing scheme for generalized seniority on spherical single-particle basis. This work redesigns the scheme to make it applicable to deformed single-particle basis. The algorithm is applied to the rare-earth nucleus $^{158}_{~64}$Gd$_{94}$ for intrinsic (body-fixed frame) neutron excitations under the low-momentum {\emph{NN}} interaction $V_{{\rm{low}}-k}$. By allowing as many as four broken pairs, we compute the lowest $300$ intrinsic states of several multipolarity. These states converge well to the exact ones, showing generalized seniority is very effective in truncating the deformed shell model. Under realistic interactions, the picture remains approximately valid that the ground state is a coherent pair condensate, and the pairs gradually break up as excitation energy increases.

%
%
%

%
%
%
%
%
%
%
\end{abstract}


\vspace{0.4in}

\maketitle

\newpage

\section{Introduction}

The nuclear shell model (configuration interaction) is the fundamental microscopic method in nuclear structure providing detailed spectroscopy. The Hamiltonian, either phenomenological or microscopic, is diagonalized in the many-body Hilbert space built as Slater determinants of single-particle levels. Nowadays the no-core shell model \cite{Barrett_2013} can treat light nuclei without assuming an inert core, in which the single-particle levels are merely a basis that affects the speed of convergence. For medium and heavy nuclei, truncation to a valence single-particle space is necessary.

Medium and heavy nuclei usually develop static deformations away from magic numbers. In general describing them in the spherical shell model is inefficient requiring large valence single-particle spaces for convergence, and successful implementations remain exceptions (see Sec. VI of Ref. \cite{Caurier_2005} and for example Refs. \cite{Alhassid_2008, Ozen_2013, Maris_2015}). Spontaneous symmetry breaking suggests using an efficient deformed single-particle basis \cite{Bohr_book, Ring_book} as first done by Nilsson \cite{Nilsson_1955}. The deformed shell model mixes Slater determinants built on the deformed valence single-particle levels and gives the intrinsic wavefunctions in the body-fixed frame.

The deformed shell model is less developed than the spherical version \cite{Caurier_2005}. Most applications are restricted to the pairing Hamiltonian. Recent advances \cite{Jensen_1995, Bogner_2003} of realistic interactions aiming at ab-initio (starting from the nucleon-nucleon potential) nuclear structure enjoy great successes with the spherical shell model \cite{Coraggio_2009}, and it is meaningful to use them in statically deformed nuclei. The plain application of the deformed shell model would suffer the same dimension problem as in the spherical case, which calls for effective truncation schemes. This work considers the generalized-seniority truncation scheme (broken pair approximation) of the deformed shell model.


In the field many competing approaches go {\emph{beyond the deformed mean field}}. We classify them by whether they conserve particle number and rotational symmetry. For methods breaking both symmetries, we mention the successful applications of the deformed quasiparticle random-phase approximation (QRPA) \cite{Peru_2011, Terasaki_2011, Nesterenko_2016} and the Hartree-Fock-Bogoliubov (HFB) plus generator coordinator method \cite{Delaroche_2010}. These applications use self-consistently the same energy-density functional for the mean-field and beyond. But they have the drawbacks of the BCS or HFB treatment that breaks the particle number \cite{Stoitsov_2007, Dobaczewski_2007} and vanishes for weak pairing \cite{Bertsch_2009}. Moreover, treating higher-order correlations better is desirable as also pointed out in these works \cite{Delaroche_2010, Terasaki_2011, Nesterenko_2016}.

For methods breaking particle number but respecting rotational symmetry, we mention the fruitful projected shell model \cite{Hara_1995, Herzberg_2006, Wang_2016}. It solves the HFB equation on Nilsson levels, then builds the basis by projecting the quasiparticle Slater determinants onto good angular momentum, on which the Hamiltonian is diagonalized. The method can be viewed as a truncation scheme of the spherical shell model. However, usually the particle-number projection is not performed thus it truncates the Fock space instead of the Hilbert space, which may arise problems (see Sec. 2.3.3. of Ref. \cite{Allaart_1988}). Also, currently the method uses phenomenological separable forces (quadrupole plus monopole and quadrupole pairing \cite{Wang_2016}) but not modern realistic ones.

For methods conserving particle number but breaking rotational symmetry, we mention the deformed shell model that diagonalizes the Hamiltonian on Slater determinants built from deformed (for example Nilsson) single-particle levels. Currently this method has been mostly applied to the (state-dependent) pairing Hamiltonian. This Hamiltonian conserves the seniority quantum number \cite{Racah_1943, Racah_1952, Flowers_1952} by which the Hilbert space is block-diagonal. Therefore the dimension is greatly reduced and the direct diagonalization is possible \cite{Molique_1997, Volya_2001}. Alternatively, the configuration-space Monte-Carlo methods \cite{Cerf_1993, Mukherjee_2011, Lingle_2015} may be more efficient, when all the pairing two-body matrix elements are attractive and free of the sign problem. We also mention that exact algebraic solutions exist for a special class \cite{Dukelsky_2004} of the pairing Hamiltonian following Richardson's method \cite{Richardson_1963_1966}. For the state-dependent pairing plus cranking Hamiltonian, Ref. \cite{Molique_1997} studied the symmetries on the many-body level that guide the truncation of the Hilbert space. For the state-independent monopole (and sometimes quadrupole) pairing plus cranking Hamiltonian, truncating the many-body basis by their energies was extensively used (for example see Refs. \cite{Zeng_1983, Zeng_1994, Xin_2000, Zhang_2011, Zhang_2012, Zhang_2013, Zhang_2013_2, Liang_2015}). Despite these achievements, the large-scale deformed shell model calculations with modern realistic interactions, comparable to those by the spherical shell model, have not been performed yet. And the effectiveness of various truncation schemes with realistic interactions remains an open question \cite{Ripoche_2017}.


For methods conserving both symmetries, we mention the deformed shell model with angular-momentum projection (projected configuration interaction method) \cite{Mishra_2008, Shukla_2009, Sahu_2013, Gao_2009_1, Gao_2009_2}. It builds the basis by projecting Slater determinants of deformed single-particle levels onto good angular momentum. The Hamiltonian is diagonalized within the basis, and the method is a truncation scheme of the spherical shell model. For better accuracy, including multiple deformations has been studied \cite{Gao_2009_2}. However, the current applications are restricted to small valence single-particle spaces (usually one spherical major shell), owing to the time-consuming angular-momentum projection. It explains the huge spherical shell model wavefunctions by using a smaller dimension, but seems not yet advancing the computation capability. In this category we also mention the MONSTER method \cite{Schmid_2004} that projects the HFB vacuum and two-quasiparticle states onto good particle number and angular momentum, on which the Hamiltonian is diagonalized.

This work concerns the deformed shell model that conserves particle number but breaks rotational symmetry. Specifically we consider the generalized-seniority truncation of it. The pairing correlation has long been recognized \cite{Bohr_1958} and influences practically all nuclei across the nuclear chart \cite{Bohr_book, Ring_book}. The generalized seniority quantum number, emphasizing pairing, was proposed \cite{Talmi_1971, Shlomo_1972, Ottaviani_1969, Gambhir_1969, Gambhir_1971} in the spherical shell model and frequently used as a truncation scheme \cite{Scholten_1983_2, Bonsignori_1985, Engel_1989, Sandulescu_1997,
Monnoye_2002, Caprio_2012_PRC,Caprio_2012_JPG, Allaart_1988, Talmi_book}. We recently proposed an algorithm \cite{Jia_2015} that greatly reduces the computer time cost and promotes the generalized-seniority truncation to an accurate tool for semi-magic nuclei \cite{Jia_2016, Qi_2016}. The concept of generalized-seniority can be straightforwardly extended to be defined on a deformed (for example Nilsson) single-particle basis. As a truncation scheme for the deformed shell model, it should be effective for the low-lying intrinsic states: deformed medium and heavy nuclei usually display pairing gaps ($\sim 1.5$ MeV) in the intrinsic spectrum. The coherent pairs are preferred by the attractive short-range pairing force, just as that in semi-magic nuclei, but are formed on the deformed single-particle levels.

Computationally, the spherical version of generalized-seniority algorithm \cite{Jia_2015} has difficulties as directly applied to the deformed case. This work redesigns the computing scheme to revive it. The generalized-seniority truncation of the deformed shell model runs as fast as that of the spherical shell model \cite{Jia_2015} with the new algorithm.

We apply the method to the rare-earth nucleus $^{158}_{~64}$Gd$_{94}$ for intrinsic (body-fixed frame) neutron excitations under the low-momentum {\emph{NN}} interaction $V_{{\rm{low}}-k}$ \cite{Bogner_2003}. The purpose is to demonstrate the effectiveness of the generalized-seniority truncation under realistic interactions. Our results show it is approximately valid that the ground state is a coherent pair condensate, and the pairs gradually break up as excitation energy increases.

The manuscript is organized as follows. Section \ref{sec_formalism} briefly reviews the generalized-seniority formalism. Section \ref{Sec_MPDM} reviews the many-pair density matrix that is key to the family of new computing schemes. We derive analytical expressions of the many-pair density matrix in Sec. \ref{Sec_MPDM2Norm}, and how this revives the spherical algorithm \cite{Jia_2015} in the case of deformed single-particle basis is explained in Sec. \ref{Sec_memory}. Section \ref{Sec_example} applies the method to the rare-earth nucleus $^{158}_{~64}$Gd$_{94}$.

\section{Generalized Seniority Formalism  \label{sec_formalism}}

We briefly review the generalized-seniority formalism in relation to the current work. For clarity we consider only one kind of nucleons, the extension to the case of active protons and neutrons is straightforward as done in for example Ref. \cite{Jia_2015}. The pair-creation operator
\begin{eqnarray}
P_\alpha^\dagger = a_\alpha^\dagger a_{\tilde{\alpha}}^\dagger  \label{P1_dag}
\end{eqnarray}
creates a pair of particles on the single-particle level $|\alpha\rangle$
and its time-reversed partner $|\tilde{\alpha}\rangle$
($|\tilde{\tilde{\alpha}}\rangle = - |\alpha\rangle$, $P_\alpha^\dagger = P_{\tilde{\alpha}}^\dagger$). The coherent
pair-creation operator
\begin{eqnarray}
P^\dagger = \sum_{m_\alpha > 0} v_\alpha P_\alpha^\dagger  \label{P_dag}
\end{eqnarray}
creates a pair of particles coherently distributed with structure
coefficients $v_\alpha$ over the entire single-particle space, where the summation runs over
orbits with a positive magnetic quantum
number $m_\alpha$. The pair-condensate wavefunction of the $2N$-particle system
\begin{eqnarray}
(P^\dagger)^{N} |{\rm vac}\rangle
 \label{gs}
\end{eqnarray}
builds in pairing correlations, where $|{\rm vac}\rangle$ is the vacuum state. The normalization is
\begin{eqnarray}
\chi_{N} = \langle{\rm vac}| P^N (P^\dagger)^{N} |{\rm vac}\rangle .  \label{chi_N}
\end{eqnarray}

Gradually breaking coherent pairs, the state with $S=2s$ unpaired nucleons is
\begin{eqnarray}
\underbrace{a^\dagger a^\dagger ... a^\dagger}_{S =
2s} (P^\dagger)^{N-s} |{\rm vac}\rangle .
\label{sen_basis}
\end{eqnarray}
Loosely speaking, $S$ is defined as the generalized-seniority quantum number \cite{Talmi_1971, Shlomo_1972, Gambhir_1969, Ottaviani_1969, Gambhir_1971, Allaart_1988, Talmi_book}. More precisely, we distinguish between the space $|S\}$ of $S$ unpaired nucleons and the space $|S\rangle$ of generalized-seniority $S$. The space $|S\}$ consists of all the states of the form (\ref{sen_basis}). Any state of $S' < S$ unpaired nucleons can be written as a linear combination of the states of $S$ unpaired nucleons, after substituting several $P^\dagger$ by Eq. (\ref{P_dag}). Therefore $|S'\}$ is a subspace of $|S\}$,
\begin{eqnarray}
| S \} \supset |S-2\}  \supset |S-4\} \supset ... \supset |2\} \supset |0\} .  \label{S_link_bp}
\end{eqnarray}
In contrast, $|S\rangle$ is the subspace after removing the subspace $|S-2\}$ from the space $|S\}$, thus
\begin{eqnarray}
&~& | S \}  \nonumber \\
&=& |S\rangle \cup | S-2 \}  \nonumber \\
&=& |S\rangle \cup | S-2 \rangle \cup |S-4\}  \nonumber \\
&=& ... \nonumber \\
&=& |S\rangle \cup |S-2\rangle \cup ... \cup |2\rangle \cup |0\rangle .  \label{le_s_space}
\end{eqnarray}
The symbol ``$\cup$'' means set union. In this work $S=2s$ is even, and we define $|s\} \equiv |S\}$ and $|s\rangle \equiv |S\rangle$. The original basis vectors (\ref{sen_basis}) are not orthogonal. After orthonormalization the new basis vectors of the space $|s\rangle$ are enumerated as $|s,i\rangle$, where the index $i$ runs from one to the dimension of $|s\rangle$.

Practical generalized-seniority calculations usually truncate the full many-body space to the subspace $|s\}$ and then diagonalize the Hamiltonian ($s = N$ corresponds to the full space without truncation). The eigen wavefunction is
\begin{eqnarray}
| E \rangle = \sum_{s' \le s} \sum_i c_{s',i} |s',i\rangle .  \label{E_wf}
\end{eqnarray}
Investigating the wavefunction (\ref{E_wf}) in terms of generalized seniority, the amplitude for generalized-seniority $2s'$ is
\begin{eqnarray}
P(s') = \sum_i |c_{s',i}|^2 .  \label{P_s}
\end{eqnarray}
And $\sum_{s' \le s} P(s') = 1$.



\section{Many-Pair Density Matrix  \label{Sec_MPDM}}

The many-pair density matrix (MPDM) \cite{Jia_2015} has clear physical meaning and is key to the new algorithm. In this section we introduce the MPDM in a natural way, and explain how it speeds up generalized-seniority calculations. We recall the conventional many-body density matrix
\begin{eqnarray}
\rho_{i_1 i_2 ... i_k;i_1' i_2' ... i_k'} \equiv \langle {\rm gs} | a_{i_1} a_{i_2} ...
a_{i_k} a_{i_1'}^\dagger a_{i_2'}^\dagger ...
a_{i_k'}^\dagger | {\rm gs} \rangle  \label{dm}
\end{eqnarray}
that characterizes properties of the ground state $| {\rm gs} \rangle$. Equation (\ref{dm}) with $k=1$ and $k=2$ give the one-body and two-body density matrix. When pairing correlation is strong, the ground state $| {\rm gs} \rangle$ can be approximated by a seniority-zero state $|{\rm gs},\nu = 0 \rangle$ (seniority $\nu$ \cite{Racah_1943, Racah_1952, Flowers_1952} and generalized seniority $S$ \cite{Talmi_1971, Shlomo_1972, Ottaviani_1969, Gambhir_1969, Gambhir_1971} are different quantum numbers), where two single-particle levels of Kramers degeneracy are either both occupied or both empty. For example we can take $|{\rm gs},\nu = 0 \rangle$ as the lowest eigenstate of diagonalizing $H$ in the $\nu = 0$ subspace. On $|{\rm gs},\nu = 0 \rangle$, the many-body density matrix $\rho_{i_1 ... i_k;i_1' ... i_k'}$ (\ref{dm}) is inefficient with many vanished matrix elements. More efficiently we introduce the MPDM
\begin{eqnarray}
t_{\alpha_1 \alpha_2 ... \alpha_p;\beta_1 \beta_2 ... \beta_p}  \nonumber \\
\equiv \langle {\rm gs},\nu = 0 | P_{\alpha_1} P_{\alpha_2} ...
P_{\alpha_p} P_{\beta_1}^\dagger P_{\beta_2}^\dagger ...
P_{\beta_p}^\dagger |{\rm gs},\nu = 0 \rangle  \label{t_ij}
\end{eqnarray}
that is physically pair-hopping amplitudes. Reference \cite{Jia_2015} shows that $\rho_{i_1 ... i_k;i_1' ... i_k'}$ of $|{\rm gs},\nu = 0 \rangle$ reduces to the form (\ref{t_ij}) in a many-to-one correspondence. Storing $t_{\alpha_1 ... \alpha_p;\beta_1 ... \beta_p}$ requires much less computer memory than storing $\rho_{i_1 ... i_k;i_1' ... i_k'}$.

The proposed fast algorithm \cite{Jia_2015} for generalized seniority has been applied to semi-magic Sn \cite{Jia_2016} and Pb \cite{Qi_2016} isotopes with realistic interactions. The key idea is to precalculate and store the MPDM. Let us consider for example computing the two-body part of the Hamiltonian. The matrix element is schematically written as
\begin{eqnarray}
\langle {\rm vac} | P^{N-s} \underbrace{a a ... a}_{S=2s} \underbrace{(a a
a^\dagger a^\dagger)}_{H} \underbrace{a^\dagger a^\dagger ...
a^\dagger}_{S=2s} (P^\dagger)^{N-s} | {\rm vac} \rangle .  \label{H_2s}
\end{eqnarray}
It is of the form of a $(S+2)$-body density matrix, and the non-vanished matrix elements reduce to the MPDM
\begin{eqnarray}
t_{\alpha_1 \alpha_2 ... \alpha_{p};\beta_1 \beta_2 ...\beta_p}^{[\gamma_1 \gamma_2 ... \gamma_r]}
= \langle {\rm vac}^{[\gamma_1 \gamma_2 ... \gamma_r]} | P^{N-s} P_{\alpha_1} P_{\alpha_2} ...
P_{\alpha_p}  \nonumber \\
\times P_{\beta_1}^\dagger P_{\beta_2}^\dagger ... P_{\beta_p}^\dagger (P^\dagger)^{N-s} | {\rm vac}^{[\gamma_1 \gamma_2 ... \gamma_r]} \rangle . ~  \label{H_t}
\end{eqnarray}
The superscripts $[\gamma_1 \gamma_2 ... \gamma_r]$ mean MPDM in the Pauli-blocked single-particle space, where pairs of single-particle levels $\gamma_1, \tilde{\gamma}_1, \gamma_2, \tilde{\gamma}_2, ... \gamma_r, \tilde{\gamma}_r$ are removed from the original single-particle space. Equation (\ref{H_t}) is the special case of Eq. (\ref{t_ij}), where the seniority-zero state $|{\rm gs},\nu = 0 \rangle$ is taken to be the generalized-seniority-zero state $|{\rm gs},S = 0 \rangle = (P^\dagger)^{N-s} | {\rm vac}^{[\gamma_1 \gamma_2 ... \gamma_r]} \rangle$.

In realistic applications usually the number of $t_{\alpha_1 \alpha_2 ... \alpha_{p};\beta_1 \beta_2 ...\beta_p}^{[\gamma_1 \gamma_2 ... \gamma_r]}$ (\ref{H_t}) is still too large to fit into memory, and further simplification is necessary. On the spherical single-particle basis, we switch to the `occupation number representation' \cite{Jia_2015, Qi_2016, Jia_2016}
\begin{eqnarray}
t_{\alpha_1 \alpha_2 ... \alpha_{p};\beta_1 \beta_2 ...\beta_p}^{[\gamma_1 \gamma_2 ... \gamma_r]} \rightarrow t_{n^\alpha_1,n^\alpha_2,...;n^\beta_1,n^\beta_2,...}^{[n^\gamma_1,n^\gamma_2,...]} ,
  \label{H_t_rot}
\end{eqnarray}
where $n^\alpha_i$ is the number of $j_i$'s (with arbitrary magnetic
quantum number $m$) present in the series
$\alpha_1,\alpha_2,...,\alpha_p$. Similarly for $n^\beta_i$ and
$n^\gamma_i$. The reduction (\ref{H_t_rot}) is justified by rotational symmetry and is again a many-to-one correspondence. As shown in Fig. 1 of Ref. \cite{Jia_2015}, $t_{n^\alpha_1,n^\alpha_2,...;n^\beta_1,n^\beta_2,...}^{[n^\gamma_1,n^\gamma_2,...]}$ could be easily stored in memory of modern computers. Precalculating $t_{n^\alpha_1,n^\alpha_2,...;n^\beta_1,n^\beta_2,...}^{[n^\gamma_1,n^\gamma_2,...]}$ is through the recursive relation (Eq. (7) of Ref. \cite{Jia_2015}).

On the deformed (for example Nilsson) single-particle basis, the reduction (\ref{H_t_rot}) is impossible in the absence of rotational symmetry. One aim of this work is to propose, in Sec. \ref{Sec_MPDM2Norm}, an alternative simplification.


\section{Express Many-pair Density Matrix by Normalization  \label{Sec_MPDM2Norm}}


Our previous works \cite{Jia_2015, Qi_2016, Jia_2016} compute MPDM through the recursive relation (Eq. (7) of Ref. \cite{Jia_2015}). In this work we propose a simpler way through expressing MPDM by the normalization (\ref{chi_N}). The new way is key to generalized-seniority on deformed single-particle basis.

We derive the results in the general case of unbalanced bra and ket generalized-seniority, and define MPDM as
\begin{eqnarray}
t_{\alpha_1 \alpha_2 ... \alpha_p;\beta_1 \beta_2 ... \beta_q}^{M}
\equiv \langle {\rm vac} | P^{M-p} P_{\alpha_1} P_{\alpha_2} ...
P_{\alpha_p}  \nonumber \\
\times P_{\beta_1}^\dagger P_{\beta_2}^\dagger ...
P_{\beta_q}^\dagger (P^\dagger)^{M-q} | {\rm vac} \rangle ,  \label{t_pq}
\end{eqnarray}
where $p$ ($q$) is the number of $\alpha$ ($\beta$) pair-indices,
and $M$ equals to the total number of pair-creation operators. Equation (\ref{H_t}) is the special case of Eq. (\ref{t_pq}) with balanced $p = q$. The $\gamma_1, \gamma_2, ..., \gamma_r$ indices are suppressed for clarity. All
the indices $\alpha_1, \alpha_2, ..., \alpha_p, \beta_1, \beta_2, ..., \beta_q$
are distinct: the MPDM vanishes if there are duplicated $\alpha$
indices, or duplicated $\beta$ indices, owing to the Pauli
principle; and we require {\emph{by definition}} that $\alpha_1, \alpha_2, ..., \alpha_p$ and $\beta_1, \beta_2, ..., \beta_q$ have no common index (the common ones act as Pauli blocking and have been moved to $\gamma_1, \gamma_2, ..., \gamma_r$).

Now we simplify Eq. (\ref{t_pq}). Substituting $P^\dagger = \sum_{m_\alpha > 0} v_\alpha P_\alpha^\dagger$ [Eq. (\ref{P_dag})] into $(P^\dagger)^{M-q}$ and polynomially expanding, terms with $P_{\beta_1}^\dagger$ vanish due to the Pauli principle. Similarly for terms with $P_{\beta_2}^\dagger$, $P_{\beta_3}^\dagger$, ..., $P_{\beta_q}^\dagger$. Thus in Eq. (\ref{t_pq}),  $(P^\dagger)^{M-q}$ could be replaced by $(P_{[\beta_1 \beta_2 ... \beta_q]}^\dagger)^{M-q}$, where
\begin{eqnarray}
P_{[\beta_1 \beta_2 ... \beta_q]}^\dagger \equiv P^\dagger - v_{\beta_1} P_{\beta_1}^\dagger - v_{\beta_2} P_{\beta_2}^\dagger - ... - v_{\beta_q} P_{\beta_q}^\dagger .  \nonumber
\end{eqnarray}
For the same reason, $P^{M-p}$ could be replaced by $(P_{[\alpha_1 \alpha_2 ... \alpha_p]})^{M-p} = (P - v_{\alpha_1} P_{\alpha_1} - ... - v_{\alpha_p} P_{\alpha_p})^{M-p}$, and Eq. (\ref{t_pq}) becomes
\begin{widetext}
\begin{eqnarray}
t_{\alpha_1 \alpha_2 ... \alpha_p;\beta_1 \beta_2 ... \beta_q}^{M}
= \langle {\rm vac} | (P_{[\alpha_1 ... \alpha_p]})^{M-p} P_{\alpha_1} P_{\alpha_2} ...
P_{\alpha_p} P_{\beta_1}^\dagger P_{\beta_2}^\dagger ...
P_{\beta_q}^\dagger (P_{[\beta_1 ... \beta_q]}^\dagger)^{M-q} | {\rm vac} \rangle .  \nonumber
\end{eqnarray}
Next, using $P_{[\alpha_1 ... \alpha_p]} = P_{[\alpha_1 ... \alpha_p \beta_1 ... \beta_q]} + v_{\beta_1} P_{\beta_1} + ... + v_{\beta_q} P_{\beta_q}$ and $P_{[\beta_1 ... \beta_q]}^\dagger = P_{[\alpha_1 ... \alpha_p \beta_1 ... \beta_q]}^\dagger + v_{\alpha_1} P_{\alpha_1}^\dagger + ... + v_{\alpha_q} P_{\alpha_q}^\dagger$, we have
\begin{eqnarray}
t_{\alpha_1 \alpha_2 ... \alpha_p;\beta_1 \beta_2 ... \beta_q}^{M}
= \langle {\rm vac} | (P_{[\alpha_1 ... \alpha_p \beta_1 ... \beta_q]} + v_{\beta_1} P_{\beta_1} + ... + v_{\beta_q} P_{\beta_q} )^{M-p} \nonumber \\
\times P_{\alpha_1} P_{\alpha_2} ...
P_{\alpha_p} P_{\beta_1}^\dagger P_{\beta_2}^\dagger ...
P_{\beta_q}^\dagger (P_{[\alpha_1 ... \alpha_p \beta_1 ... \beta_q]}^\dagger + v_{\alpha_1} P_{\alpha_1}^\dagger + ... + v_{\alpha_p} P_{\alpha_p}^\dagger )^{M-q} | {\rm vac} \rangle .  \nonumber
\end{eqnarray}
In the polynomial expansion of $(P_{[\alpha_1 ... \alpha_p \beta_1 ... \beta_q]} + v_{\beta_1} P_{\beta_1} + ... + v_{\beta_q} P_{\beta_q} )^{M-p}$, each contributing term must have the factor $P_{\beta_1} P_{\beta_2} ...
P_{\beta_q}$ to annihilate $P_{\beta_1}^\dagger P_{\beta_2}^\dagger ... P_{\beta_q}^\dagger$. Defining $A_a^b = a!/(a-b)!$ as the number of permutations, we write $$(P_{[\alpha_1 ... \alpha_p \beta_1 ... \beta_q]} + v_{\beta_1} P_{\beta_1} + ... + v_{\beta_q} P_{\beta_q} )^{M-p} =(P_{[\alpha_1 ... \alpha_p \beta_1 ... \beta_q]} )^{M-p-q} A_{M-p}^{q} v_{\beta_1} P_{\beta_1} v_{\beta_2} P_{\beta_2} ... v_{\beta_q} P_{\beta_q} + ...$$
The neglected terms `...' do not contribute. Treating $(P_{[\alpha_1 ... \alpha_p \beta_1 ... \beta_q]}^\dagger + v_{\alpha_1} P_{\alpha_1}^\dagger + ... + v_{\alpha_p} P_{\alpha_p}^\dagger )^{M-q}$ similarly, Eq. (\ref{t_pq}) becomes
\begin{eqnarray}
t_{\alpha_1 \alpha_2 ... \alpha_p;\beta_1 \beta_2 ... \beta_q}^{M}
= \langle {\rm vac} | (P_{[\alpha_1 ... \alpha_p \beta_1 ... \beta_q]} )^{M-p-q} A_{M-p}^{q} v_{\beta_1} P_{\beta_1} ... v_{\beta_q} P_{\beta_q} \nonumber \\
\times P_{\alpha_1} P_{\alpha_2} ...
P_{\alpha_p} P_{\beta_1}^\dagger P_{\beta_2}^\dagger ...
P_{\beta_q}^\dagger A_{M-q}^{p} v_{\alpha_1} P_{\alpha_1}^\dagger ... v_{\alpha_p} P_{\alpha_p}^\dagger (P_{[\alpha_1 ... \alpha_p \beta_1 ... \beta_q]}^\dagger )^{M-p-q} | {\rm vac} \rangle  \nonumber \\
= A_{M-q}^{p} A_{M-p}^{q} v_{\alpha_1} ... v_{\alpha_p} v_{\beta_1} ... v_{\beta_q} \langle {\rm vac} | (P_{[\alpha_1 ... \alpha_p \beta_1 ... \beta_q]} )^{M-p-q}(P_{[\alpha_1 ... \alpha_p \beta_1 ... \beta_q]}^\dagger )^{M-p-q} | {\rm vac} \rangle .  \nonumber
\end{eqnarray}
Defining $\chi_{M-p-q}^{[\alpha_1 ... \alpha_p \beta_1 ... \beta_q]} = \langle {\rm vac} | (P_{[\alpha_1 ... \alpha_p \beta_1 ... \beta_q]} )^{M-p-q}(P_{[\alpha_1 ... \alpha_p \beta_1 ... \beta_q]}^\dagger )^{M-p-q} | {\rm vac} \rangle$ as the normalization (\ref{chi_N}) in the Pauli-blocked single-particle space, and using $A_{M-q}^{p} A_{M-p}^{q}
= (M-p)!(M-q)!/[(M-p-q)!]^2$, we have
\begin{eqnarray}
t_{\alpha_1 \alpha_2 ... \alpha_p;\beta_1 \beta_2 ... \beta_q}^{M}
= \frac{(M-p)!(M-q)!}{[(M-p-q)!]^2} v_{\alpha_1} v_{\alpha_2} ... v_{\alpha_p} v_{\beta_1} v_{\beta_2} ... v_{\beta_q} \chi_{M-p-q}^{[\alpha_1 \alpha_2 ... \alpha_p \beta_1 \beta_2 ... \beta_q]} .
  \label{t_pq_res}
\end{eqnarray}
This finishes the derivation.

In this work the relevant MPDM (\ref{H_t}) has balanced bra and ket generalized seniority. Equation (\ref{t_pq}) becomes Eq. (\ref{H_t}) after setting $q = p$ and $M = N-s+p$. The derivation from Eq. (\ref{t_pq}) to Eq. (\ref{t_pq_res}) remains valid if we Pauli block the $\gamma_1, \gamma_2, ..., \gamma_r$ indices from the very beginning. Therefore Eq. (\ref{t_pq_res}), with these settings, implies the result for Eq. (\ref{H_t})
\begin{eqnarray}
t_{\alpha_1 \alpha_2 ... \alpha_p;\beta_1 \beta_2 ... \beta_p}^{[\gamma_1 \gamma_2 ... \gamma_r]}  = \langle {\rm vac}^{[\gamma_1 \gamma_2 ... \gamma_r]} | P^{N-s} P_{\alpha_1} P_{\alpha_2} ...
P_{\alpha_p} P_{\beta_1}^\dagger P_{\beta_2}^\dagger ...
P_{\beta_p}^\dagger (P^\dagger)^{N-s} | {\rm vac}^{[\gamma_1 \gamma_2 ... \gamma_r]} \rangle  \nonumber \\
= [ \frac{(N - s)!}{(N-s-p)!} ]^2 v_{\alpha_1} v_{\alpha_2} ... v_{\alpha_p} v_{\beta_1} v_{\beta_2} ... v_{\beta_p} \chi_{N-s-p}^{[\alpha_1 \alpha_2 ... \alpha_p \beta_1 \beta_2 ... \beta_p \gamma_1 \gamma_2 ... \gamma_r]} .
  \label{t_pp_res}
\end{eqnarray}
\end{widetext}
Equation (\ref{t_pp_res}) expresses MPDM by the Pauli-blocked normalizations in a many-to-one correspondence. This result is key to generalized seniority on deformed single-particle basis as will be shown in Sec. \ref{Sec_memory}.

\section{Estimate Computer Memory \label{Sec_memory}}


The family of new algorithms speeds up generalized-seniority calculations by precalculating and storing in memory the selected intermediate quantity. In this section we compare the memory requirements of the two methods by selecting $t_{\alpha_1 \alpha_2 ... \alpha_p;\beta_1 \beta_2 ... \beta_p}^{[\gamma_1 \gamma_2 ... \gamma_r]}$ [Eq. (\ref{H_t}) or the left-hand side of Eq. (\ref{t_pp_res})] and by selecting $\chi_{N-s-p}^{[\alpha_1 ... \alpha_p \beta_1 ... \beta_p \gamma_1 ... \gamma_r]}$ [the right-hand side of Eq. (\ref{t_pp_res})] as the intermediate quantity.

Taking the two-body part of the Hamiltonian as an example and in the reduction from Eq. (\ref{H_2s}) to Eq. (\ref{H_t}), three restrictions exist on the value of
non-negative integers $r$ and $p$ ($2\Omega$ is the dimension of the single-particle space),
\begin{eqnarray}
r + 2p \le 2(s+1) \le 2r + 2p ,  \label{con1_rp} \\
p + (N-s) + r \le \Omega ,  \label{con2_rp} \\
p \le N-s .  \label{con3_rp}
\end{eqnarray}
Given $\Omega$, $N$, and $s$, these three equations determine the possible values of the $(p,r)$ pair. For each $(p,r)$ pair, the number of different $t_{\alpha_1 \alpha_2 ... \alpha_p;\beta_1 \beta_2 ... \beta_p}^{[\gamma_1 \gamma_2 ... \gamma_r]}$ is
\begin{eqnarray}
k_t(p,r) = \frac{1}{2} C_{\Omega}^r C_{\Omega-r}^p C_{\Omega-r-p}^p = \frac{1}{2} \frac{\Omega!}{r!p!p!(\Omega-r-2p)!} .  \nonumber
\end{eqnarray}
$C_{\Omega}^r = \Omega!/[r!(\Omega-r)!]$ is the number of ways for selecting $r$ of $\gamma$ indices from the $\Omega$ candidates that compose the single-particle space. $C_{\Omega-r}^p$ is for selecting $p$ of $\alpha$ indices from the leftover $\Omega-r$ candidates. Similarly $C_{\Omega-r-p}^p$ is for selecting $p$ of $\beta$ indices. The factor $1/2$ considers that $t_{\alpha_1 \alpha_2 ... \alpha_p;\beta_1 \beta_2 ... \beta_p}^{[\gamma_1 \gamma_2 ... \gamma_r]} = t_{\beta_1 \beta_2 ... \beta_p;\alpha_1 \alpha_2 ... \alpha_p}^{[\gamma_1 \gamma_2 ... \gamma_r]}$ is symmetric exchanging the $\alpha$ and $\beta$ indices. $\sum_{(p,r)} k_t(p,r)$ sums all possible $(p,r)$ pairs and gives the total number of different $t_{\alpha_1 \alpha_2 ... \alpha_p;\beta_1 \beta_2 ... \beta_p}^{[\gamma_1 \gamma_2 ... \gamma_r]}$ at given $\Omega$, $N$, and $s$. Next, for each $(p,r)$ pair, the number of different $\chi_{N-s-p}^{[\alpha_1 ... \alpha_p \beta_1 ... \beta_p \gamma_1 ... \gamma_r]}$ is
\begin{eqnarray}
k_{\chi}(p,r) = C_{\Omega}^{2p+r} = \frac{\Omega!}{(r+2p)!(\Omega-r-2p)!} .
\end{eqnarray}
$C_{\Omega}^{2p+r}$ is the number of ways for selecting $2p+r$ indices $\alpha_1 ... \alpha_p \beta_1 ... \beta_p \gamma_1 ... \gamma_r$ from the $\Omega$ candidates. $\sum_{(p,r)} k_{\chi}(p,r)$ sums all possible $(p,r)$ pairs and gives the total number of different $\chi_{N-s-p}^{[\alpha_1 ... \alpha_p \beta_1 ... \beta_p \gamma_1 ... \gamma_r]}$ at given $\Omega$, $N$, and $s$.

In for example {\emph{Matlab}}, each double-precision variable occupies $64$ bits or $4$ bytes memory. The complete $t$ table needs $4$ variables for each $t_{\alpha_1 \alpha_2 ... \alpha_p;\beta_1 \beta_2 ... \beta_p}^{[\gamma_1 \gamma_2 ... \gamma_r]}$, storing not only the value, but also the $3$ indices $\alpha_1 ... \alpha_p$, $\beta_1 ... \beta_p$, and $\gamma_1 ... \gamma_r$ (each in one variable bitwise). Hence the $t$ table needs $M_t(\Omega,N,s) = 16 \sum_{(p,r)} k_{t}(p,r)$ bytes memory. The complete $\chi$ table needs $2$ variables for each $\chi_{N-s-p}^{[\alpha_1 ... \alpha_p \beta_1 ... \beta_p \gamma_1 ... \gamma_r]}$, storing the value and the $\alpha_1 ... \alpha_p \beta_1 ... \beta_p \gamma_1 ... \gamma_r$ index. (The $p$ index can be stored in an $8$-bit variable and used in a first-level indexing, thus the memory cost is small and neglected.) Hence the $\chi$ table needs $M_{\chi}(\Omega,N,s) = 8 \sum_{(p,r)} k_{\chi}(p,r)$ bytes memory.

\begin{figure}
\includegraphics[width = 0.5\textwidth]{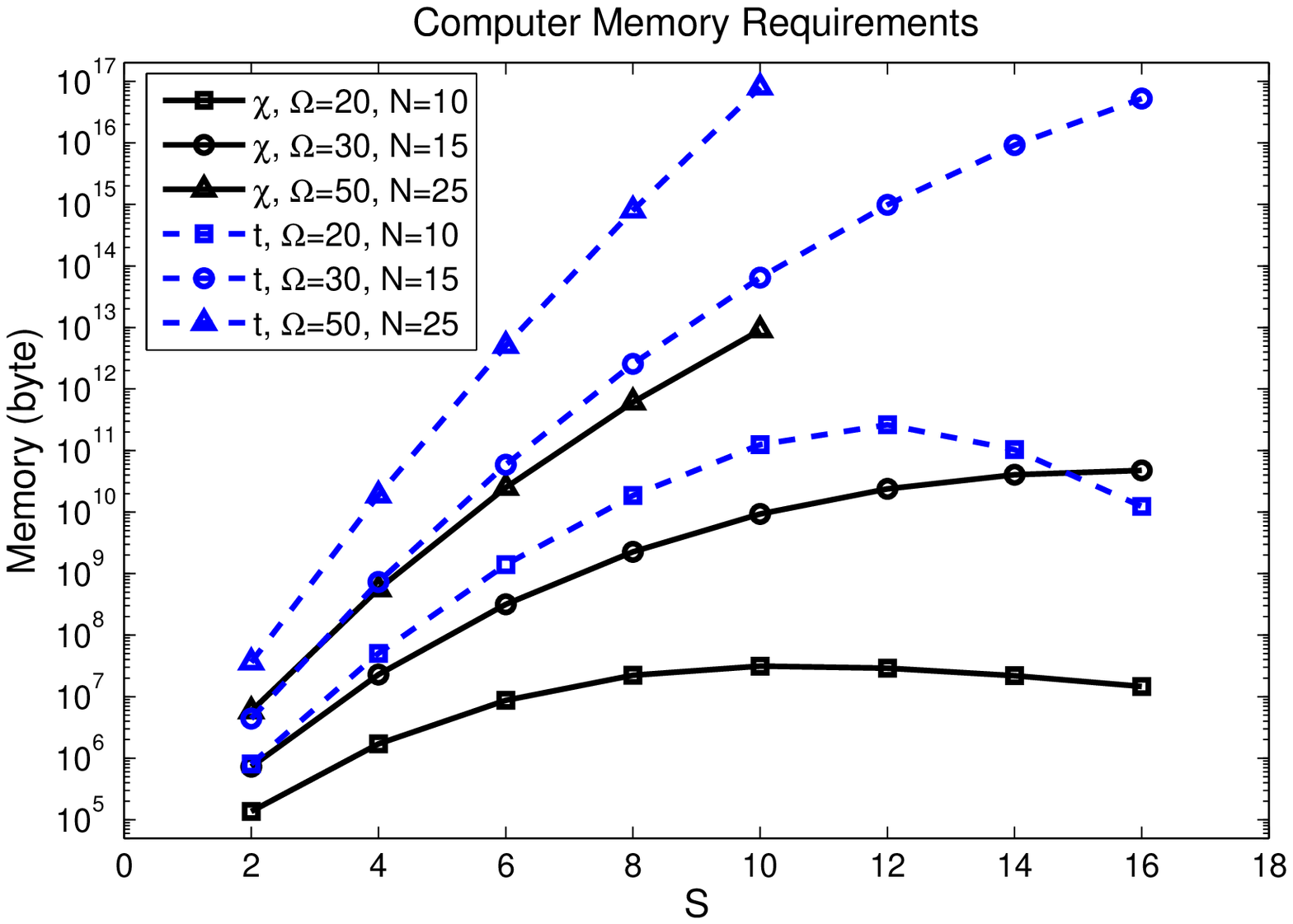}
\caption{\label{Fig_memory} (Color online) Memory requirements to store the $\chi$ and $t$ tables in three different model spaces. $2 \Omega$ is the dimension of the single-particle space, $2 N$ is the number of particles, and $S$ is the generalized seniority quantum number. }
\end{figure}

Figure \ref{Fig_memory} plots the memory requirements for the $t$ table [$M_{t}(\Omega,N,s)$] and the $\chi$ table [$M_{\chi}(\Omega,N,s)$] in three model spaces of $(\Omega,N) = (20,10)$, $(30,15)$, and $(50,25)$. In each case the space is half filled with $N = \Omega/2$. We see that the $\chi$ table is considerably smaller than the $t$ table. For large model spaces, it is impractical or difficult to store the $t$ table in memory of common modern computers (several dozens of GB, $1$GB $\approx 10^9$ bytes), especially if parallel computing stores multiple copies of the $t$ table. Even if memory is enough, it is preferable to use a small table that is constantly searched in the algorithm.

To summarize, in realistic deformed applications the number of $t_{\alpha_1 \alpha_2 ... \alpha_p;\beta_1 \beta_2 ... \beta_p}^{[\gamma_1 \gamma_2 ... \gamma_r]}$ is frequently too large to fit into memory. In this work the proposed deformed generalized-seniority algorithm precalculates (by Eq. (23) of Ref. \cite{Jia_2013_1}) and stores the Pauli-blocked normalizations $\chi_{N-s-p}^{[\alpha_1 ... \alpha_p \beta_1 ... \beta_p \gamma_1 ... \gamma_r]}$, then computes $t_{\alpha_1 \alpha_2 ... \alpha_p;\beta_1 \beta_2 ... \beta_p}^{[\gamma_1 \gamma_2 ... \gamma_r]}$ on the fly through Eq. (\ref{t_pp_res}). Section \ref{Sec_example} applies the algorithm to a semi-realistic example.


\section{Semi-realistic Example  \label{Sec_example}}

In this section we apply the generalized seniority approximation to the semi-realistic example of the rare-earth nucleus $^{158}_{~64}$Gd$_{94}$. The purpose is to demonstrate the effectiveness of this truncation scheme under realistic interactions. For simplicity, we consider only the neutron degree of freedom, governed by the anti-symmetrized two-body Hamiltonian
\begin{eqnarray}
H = \sum_{1} e_1 a_1^\dagger a_1 + \frac{1}{4} \sum_{1234} V_{1234} a_1^\dagger a_2^\dagger a_3 a_4 . \label{H_def}
\end{eqnarray}
The single-particle levels $e_1$ are assumed to be the eigenstates of the Nilsson model \cite{Nilsson_1955},
\begin{eqnarray}
h = - \frac{\hbar^2}{2m} \nabla^2 + \frac{m}{2} (\omega_r^2 x^2 + \omega_r^2 y^2 + \omega_z^2 z^2)  \nonumber \\
- \kappa \hbar \mathring{\omega}_0 [ 2 \textbf{l} \cdot \textbf{s} + \mu ( \textbf{l}^2 - \langle \textbf{l}^2 \rangle_{\mathcal{N}} ) ] ,     \label{H_Nil}
\end{eqnarray}
where $\hbar \mathring\omega_0 = 41 A^{-1/3} {\rm {MeV}}$ as usual and $A = 158$ is the mass number. Other parameters follow the convention in Ref. \cite{Ring_book}. Taking the experimental quadrupole deformation $\beta = \frac{4}{3} \sqrt{\frac{\pi}{5}} \delta = 1.0569 \delta = 0.349$ \cite{NNDC}, $\omega_r$ and $\omega_z$ are fixed by $2\delta = 3(\omega_r^2 - \omega_z^2)/(2 \omega_r^2 + \omega_z^2)$ and conserving the volume $(\omega_r)^2 \omega_z = (\mathring{\omega}_0)^3$. $\langle \textbf{l}^2 \rangle_{\mathcal{N}} = \mathcal{N} (\mathcal{N}+3) / 2$ is $\textbf{l}^2$ averaged over one harmonic oscillator major shell $\mathcal{N} = 2n_r+l$. We take $\kappa = 0.0637$ and $\mu = 0.60$ as commonly used \cite{Gustafson_1967, Ring_book}.


The neutron residual interaction $V_{1234}$ in Eq. (\ref{H_def}) is assumed to be the low-momentum {\emph{NN}} interaction $V_{{\rm{low}}-k}$ \cite{Bogner_2003} derived from the free-space N$^3$LO potential \cite{Entem_2003}. Practically, we use the code distributed by M. Hjorth-Jensen \cite{Morten_code} to compute (without Coulomb, charge-symmetry breaking, or charge-independence breaking) the two-body matrix elements of $V_{{\rm{low}}-k}$ in the spherical harmonic oscillator basis up to (including) the $\mathcal{N} = 12$ major shell, with the standard momentum cutoff $2.1$ fm$^{-1}$. The Nilsson model (\ref{H_Nil}) is diagonalized in this spherical $\mathcal{N} \le 12$ basis, the eigen energies are $e_1$ and the eigen wavefunctions transform the spherical two-body matrix elements into those on the Nilsson basis as used in the Hamiltonian (\ref{H_def}).


The above procedure assumes that mainly the proton-neutron interaction generates the static deformation and self-consistently the Nilsson mean-field. The residual proton-neutron interaction is neglected, and in the Hamiltonian (\ref{H_def}) the part of the neutron-neutron interaction already included in the Nilsson mean field $e_1$ is not removed from $V_{1234}$. These assumptions make the example semi-realistic. Our goal is to demonstrate the effectiveness of the generalized-seniority truncation scheme, not to accurately reproduce the experimental data.

\begin{figure}
\includegraphics[width = 0.5\textwidth]{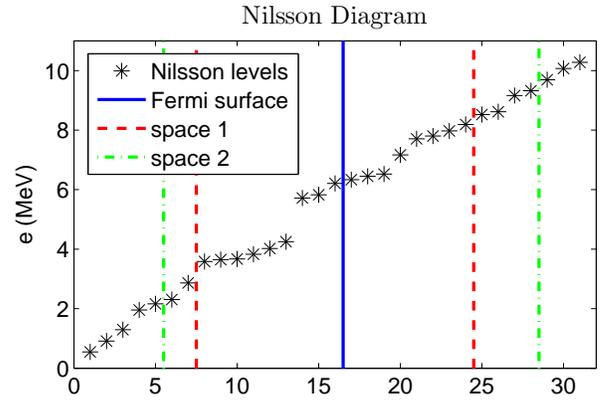}
\caption{\label{Fig_Nilsson_diagram} (Color online) Part of the Nilsson diagram. The $31$ asterisks represent $31$ consecutive pairs of Nilsson levels. The horizontal and vertical axes show their numbering and energy (zero of energy is arbitrary). The blue solid line is the Fermi surface. The Nilsson levels between the two red dashed (green dash-dot) lines compose the valence space $1$ ($2$). }
\end{figure}

The Fermi energy is fixed as usual to be the average of the last occupied and the first unoccupied Nilsson level (when the $94$ neutrons occupy the lowest $47$ pairs of Nilsson levels). We perform two calculations in two valence single-particle spaces of dimension $34$ and $46$ as shown in Fig. \ref{Fig_Nilsson_diagram}. The dimension-$34$ space has $18$ and $16$ valence levels below and above the Fermi surface, and in calculation $1$ we truncate the many-body space up to generalized-seniority $S = 8$. The dimension-$46$ space has $22$ and $24$ valence levels below and above the Fermi surface, and in calculation $2$ we truncate up to $S = 6$. In each calculation, the pair structure $v_\alpha$ (\ref{P_dag}) is determined by the variation principle (using Matlab function `{\emph{fminunc}}'). The Hamiltonian (\ref{H_def}) conserves parity $\pi$ and angular-momentum projection $K$ onto the intrinsic symmetry axis. We compute in the Lanczos method the lowest $300$ eigenstates for $K = 0, 2, 3, 6, 10$ and both parities.

\begin{figure}
\includegraphics[width = 0.45\textwidth]{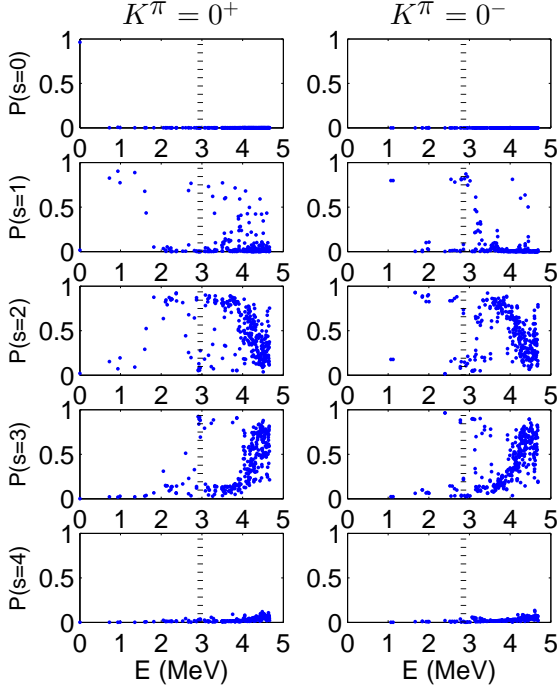}
\caption{\label{Fig_N9_s4_K0_Ps} (Color online)  Amplitudes $P(s)$ of each generalized seniority $S=2s$ versus the excitation energy of the $K = 0$ eigenstates by calculation 1. The left (right) panels plot the lowest $300$ eigenstates with positive (negative) parity. Therefore each panel has $300$ data points. The vertical dotted line is $E = E_{s=1, K^\pi = 0^+}^{<}$ for the left panels and $E = E_{s=1, K^\pi = 0^-}^{<}$ for the right panels. }
\end{figure}
\begin{figure}
\includegraphics[width = 0.45\textwidth]{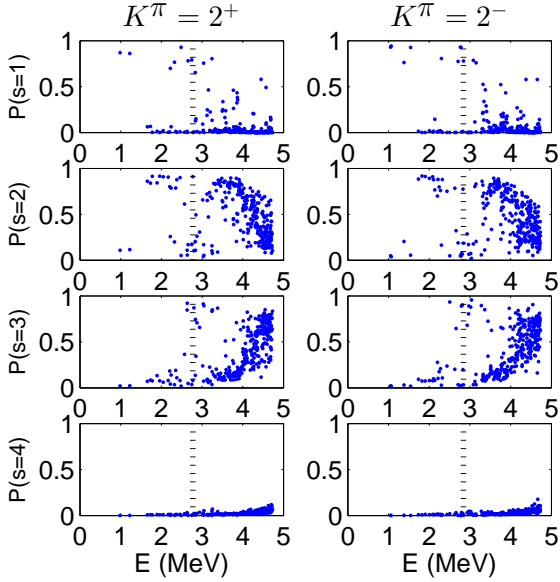}
\caption{\label{Fig_N9_s4_K2_Ps} (Color online)  Amplitudes of each generalized seniority versus the excitation energy of the $K = 2$ eigenstates by calculation 1. The $P(s=0)$ amplitudes vanish owing to symmetry, and are not plotted. }
\end{figure}
\begin{figure}
\includegraphics[width = 0.45\textwidth]{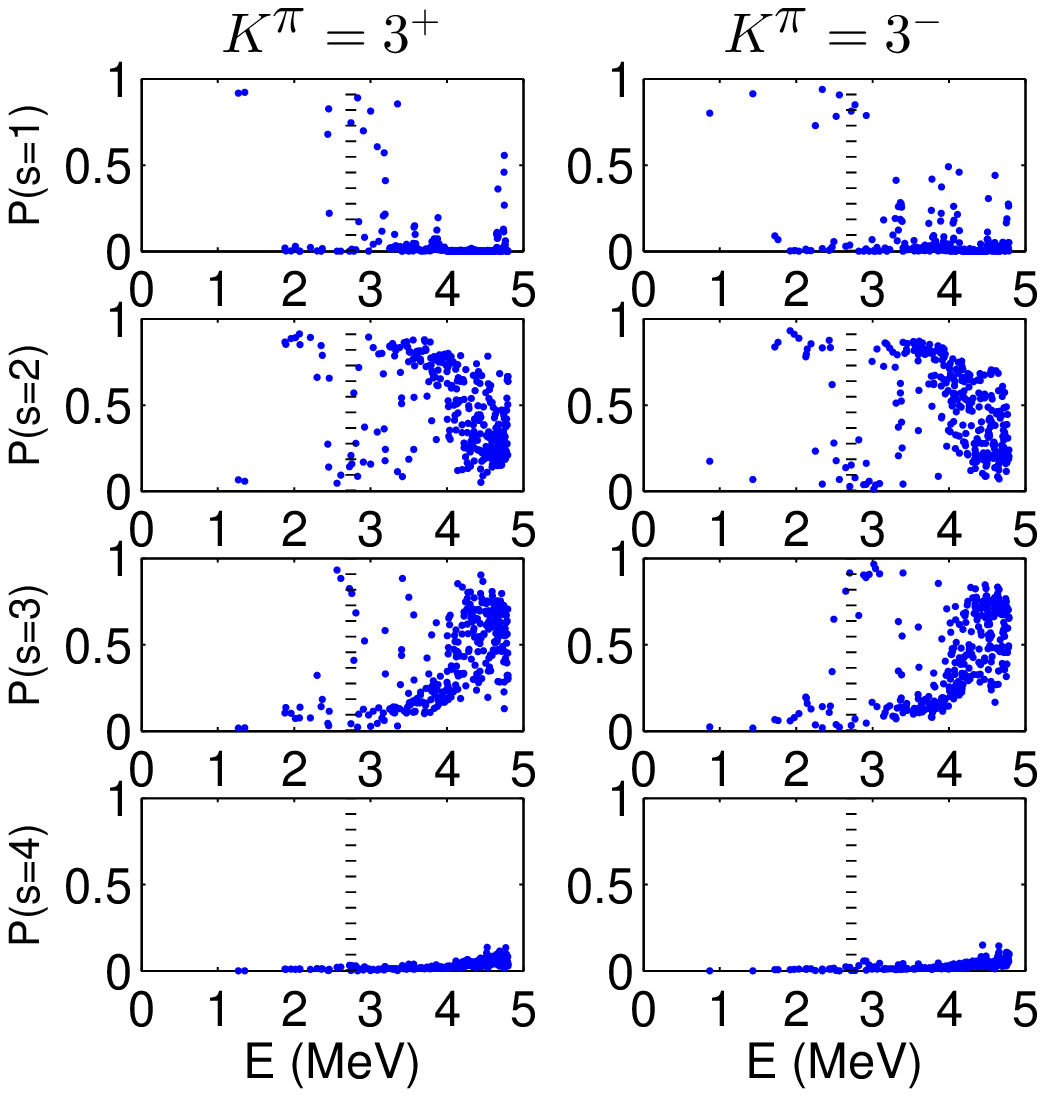}
\caption{\label{Fig_N9_s4_K3_Ps} (Color online)  Amplitudes of each generalized seniority versus the excitation energy of the $K = 3$ eigenstates by calculation 1. }
\end{figure}
\begin{figure}
\includegraphics[width = 0.45\textwidth]{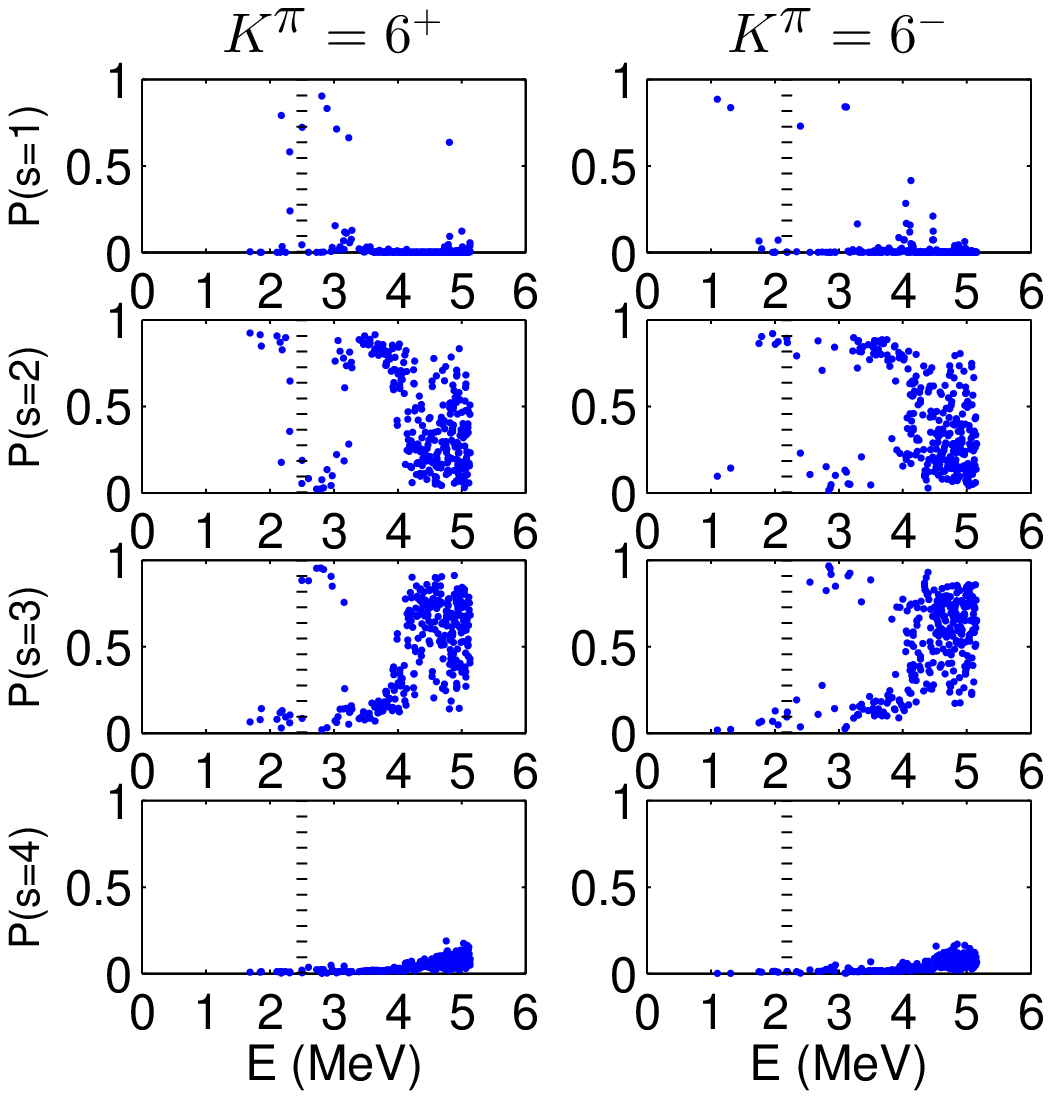}
\caption{\label{Fig_N9_s4_K6_Ps} (Color online)  Amplitudes of each generalized seniority versus the excitation energy of the $K = 6$ eigenstates by calculation 1. }
\end{figure}
\begin{figure}
\includegraphics[width = 0.45\textwidth]{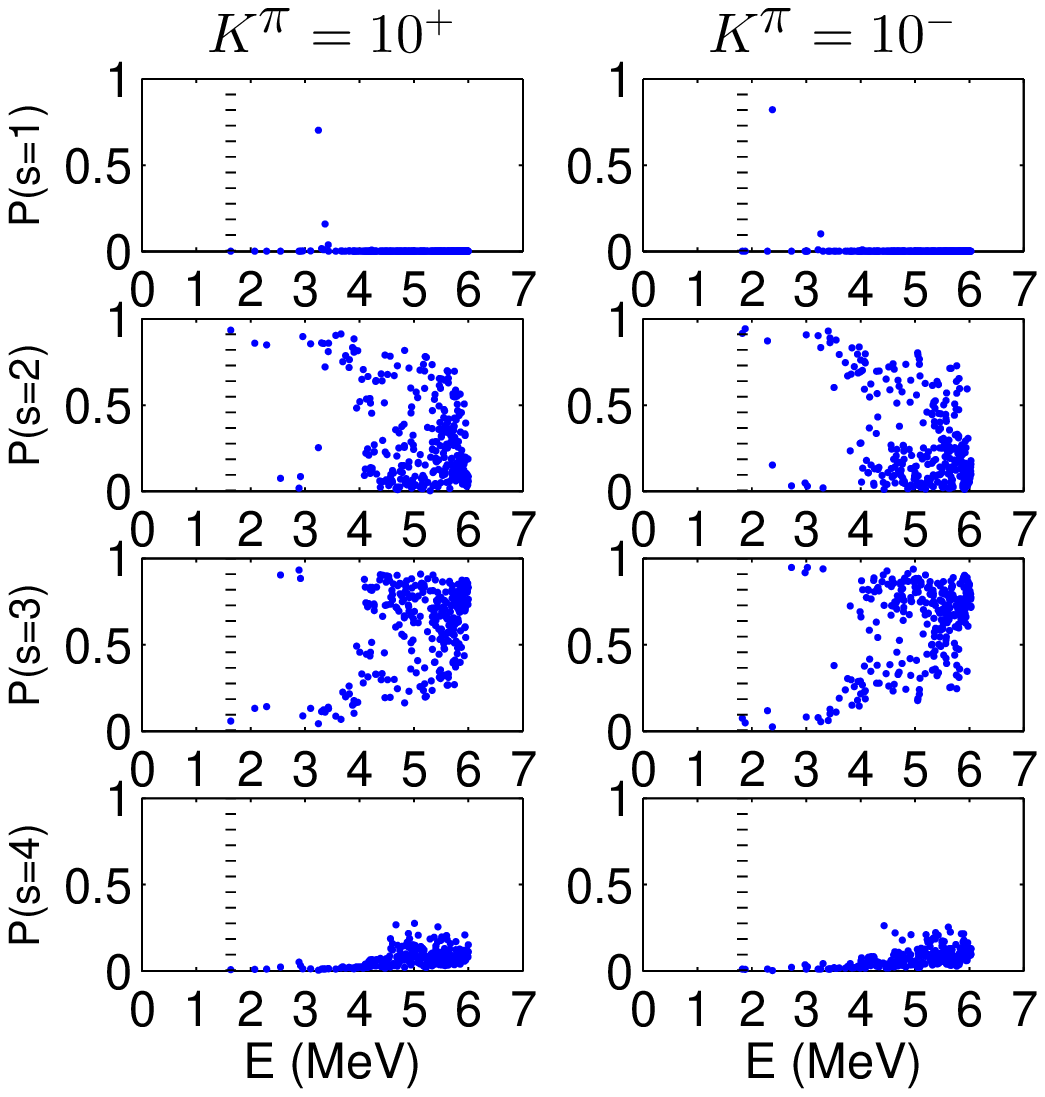}
\caption{\label{Fig_N9_s4_K10_Ps} (Color online)  Amplitudes of each generalized seniority versus the excitation energy of the $K = 10$ eigenstates by calculation 1. }
\end{figure}

Figures \ref{Fig_N9_s4_K0_Ps}-\ref{Fig_N9_s4_K10_Ps} show the results of calculation $1$. The vertical axes show the amplitudes $P(s)$ (\ref{P_s}), and the horizontal axes show the excitation energies $E$. If the pairing force was very strong, the states with $s$ broken-pairs were roughly degenerate at $s$ times the pairing gap. In reality other correlations and the non-degeneracy of Nilsson levels disturb this picture. The pairing gap is about $1.5$ MeV for deformed medium and heavy nuclei, smaller than the gap of spherical semi-magic nuclei around $2$ MeV. We would ask whether the ground state is still a condensate of coherent pairs (\ref{gs}), to what extent the condensed pairs gradually break up as the excitation energy increases, and if the generalized-seniority truncation remains effective.

The left panels of Fig. \ref{Fig_N9_s4_K0_Ps} show that the ground state (the point at $E = 0$) is a very good pair condensate. The $P(s=0)$ component (\ref{gs}) dominates the wavefunction, the $P(s=1)$ and $P(s=2)$ amplitudes are tiny, and the $P(s \ge 3)$ amplitudes are negligible. This suggests that the variation principle on the trial wavefunction (\ref{gs}), conventionally called variation after particle-number projection, may be a very accurate method for deformed nuclei with realistic interactions. To further improve the wavefunction, it may be enough to include up to generalized-seniority $S = 2s = 4$ (recall the QRPA ground state is the quasiparticle vacuum mixed mainly with the four-quasiparticle components). This could be easily done by the new algorithm in large valence spaces with over $100$ Nilsson levels.

In Figs. \ref{Fig_N9_s4_K0_Ps}-\ref{Fig_N9_s4_K10_Ps} the pattern is recognizable that the condensed pairs gradually break up as the excitation energy increases, but strong mixing among different $S$ exists in the wavefunctions. Also there are many examples of high $S$ states intruding into low energies. As an indicator, we introduce the symbol $E_{s=1, K^\pi = 0^+}^{<}$ as the energy below which the number of many-body eigenstates is equal to the dimension of the $|s=1, K^\pi = 0^+\}$ subspace (\ref{S_link_bp}). The vertical dotted lines on these figures represent $E = E_{s=1,K^\pi}^{<}$ for different $K^\pi$, to the left of this line the number of data points is equal to the dimension of the $|s=1, K^\pi\}$ subspace. We find many states intruding to the left of this line have large $P(s=2)$ and moderate $P(s=3)$ amplitudes. Similar figures have been plotted for semi-magic Sn isotopes (Figs. 13-21 and 23-26 of Ref. \cite{Jia_2016}). Comparing with those figures, the pattern of gradual breakup of condensed pairs is less obvious in deformed nuclei than in semi-magic nuclei.

The method truncates the shell-model space to $|S\}$ (\ref{S_link_bp}). Increasing $S$ and thus the subspace size, the eigen wavefunctions gradually converge to the exact shell-model
ones when all the pairs are broken ($S = 2s = 2N$). The truncation scheme is effective if it converges fast. Figures \ref{Fig_N9_s4_K0_Ps}-\ref{Fig_N9_s4_K10_Ps} show that the $P(s=4)$ amplitudes are small, especially below $4$ MeV in excitation energy. No exception exists, therefore we should not miss any shell-model eigenstate. This indicates that the wavefunctions have converged very well, and the generalized-seniority truncation is effective. The dimension of the $|S=8, K^\pi\}$ subspace is approximately $522$, $502$, $477$, $364$, $187$ thousand for $K = 0, 2, 3, 6, 10$ and positive parity. The dimensions for negative parity are approximately the same.

\begin{figure}
\includegraphics[width = 0.45\textwidth]{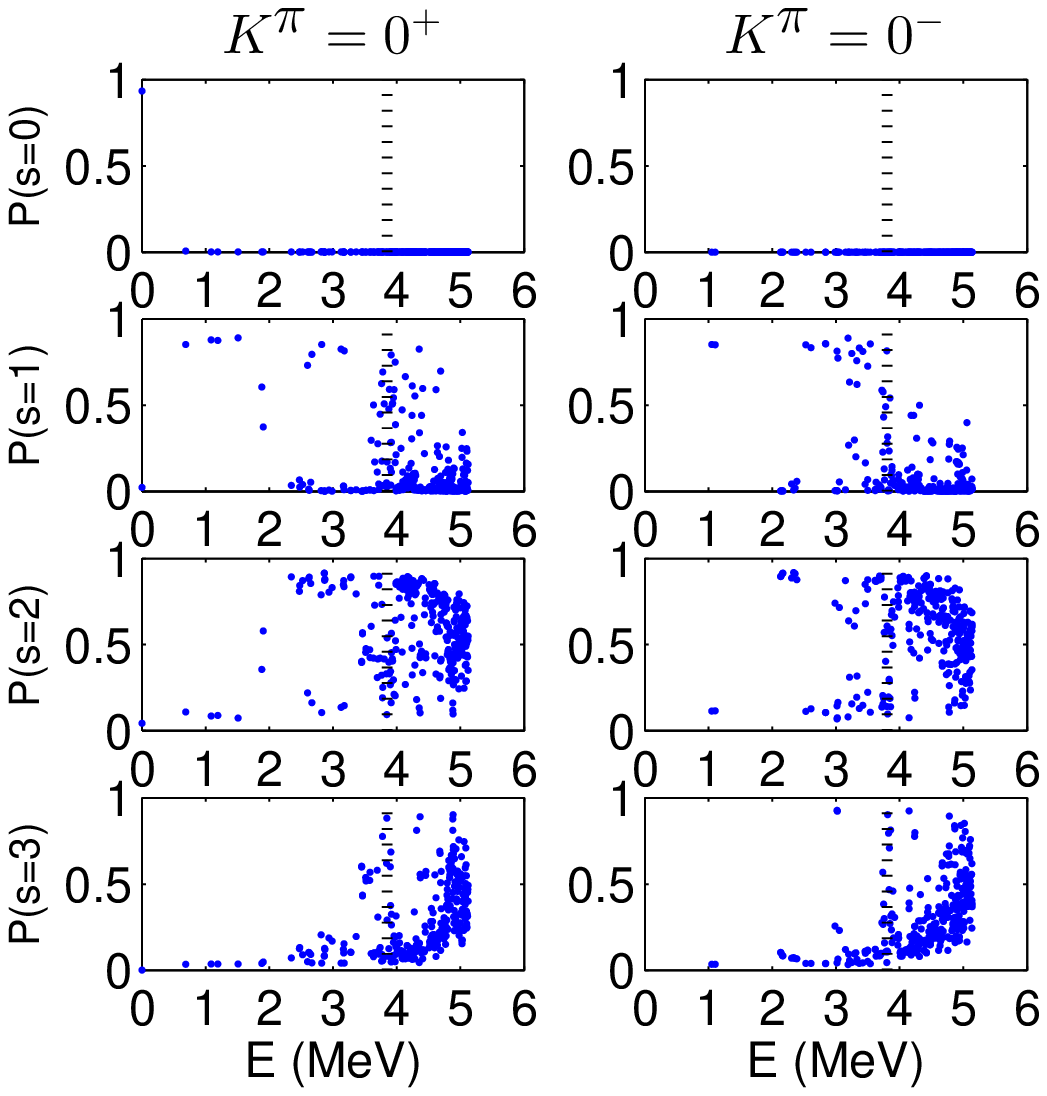}
\caption{\label{Fig_N11_s3_K0_Ps} (Color online)  Amplitudes of each generalized seniority versus the excitation energy of the $K = 0$ eigenstates by calculation 2. }
\end{figure}

\begin{figure}
\includegraphics[width = 0.45\textwidth]{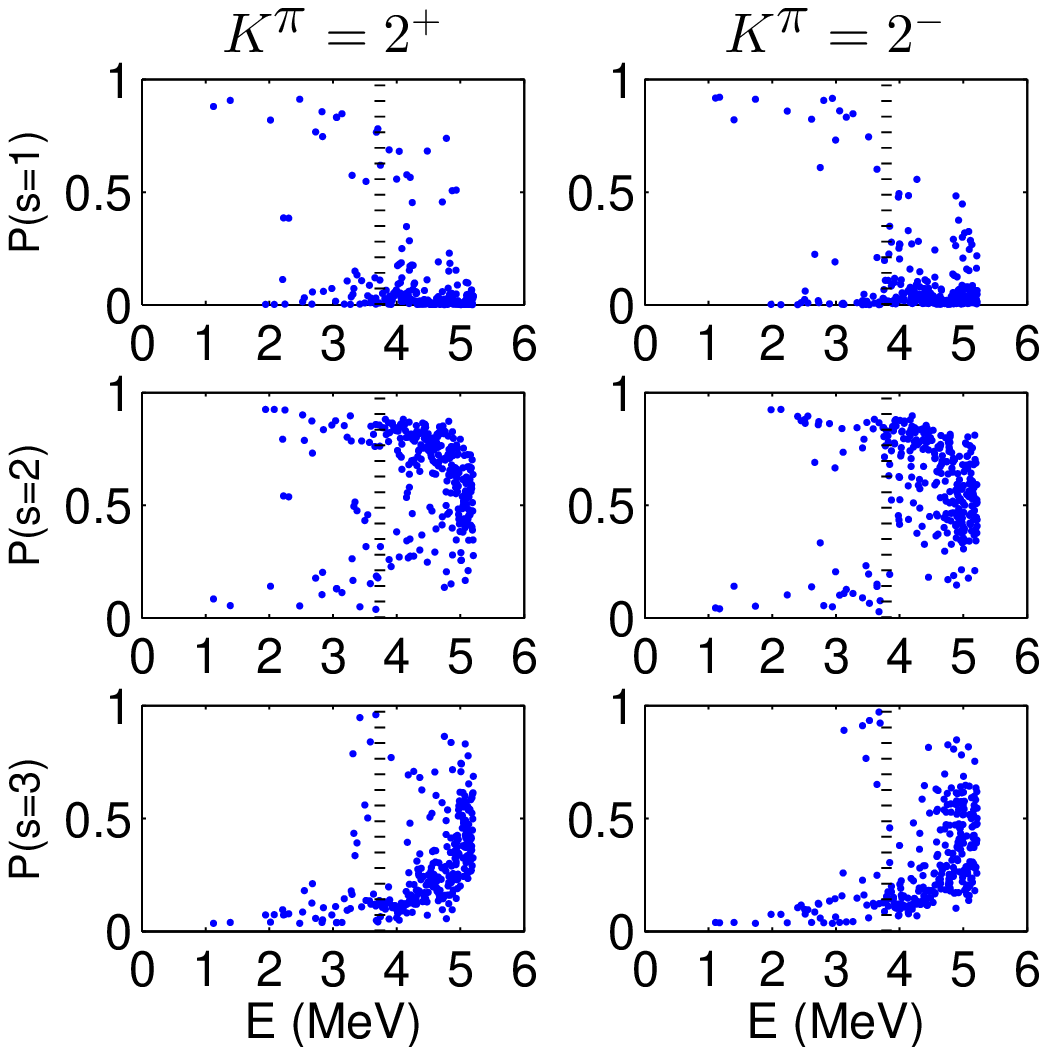}
\caption{\label{Fig_N11_s3_K2_Ps} (Color online)  Amplitudes of each generalized seniority versus the excitation energy of the $K = 2$ eigenstates by calculation 2. }
\end{figure}

\begin{figure}
\includegraphics[width = 0.45\textwidth]{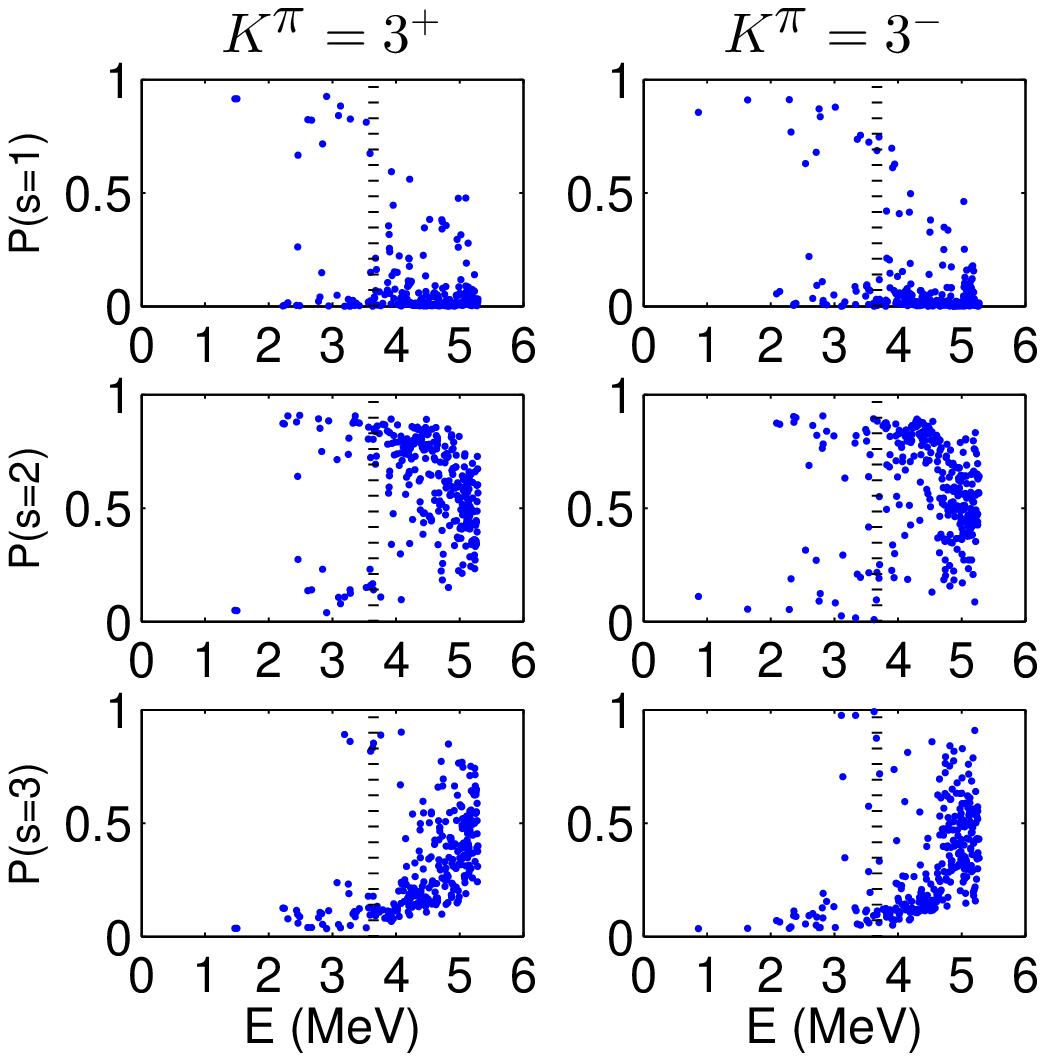}
\caption{\label{Fig_N11_s3_K3_Ps} (Color online)  Amplitudes of each generalized seniority versus the excitation energy of the $K = 3$ eigenstates by calculation 2. }
\end{figure}

\begin{figure}
\includegraphics[width = 0.45\textwidth]{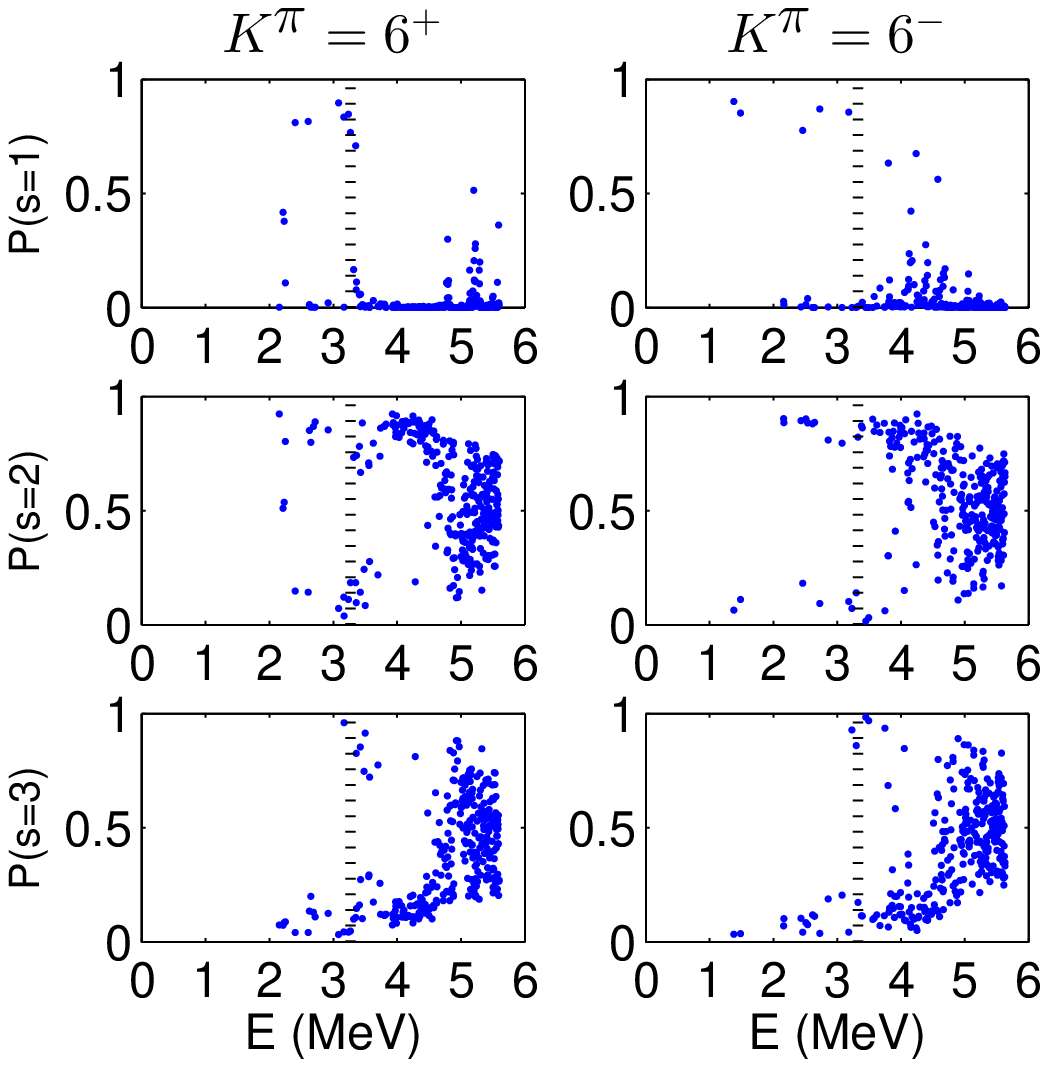}
\caption{\label{Fig_N11_s3_K6_Ps} (Color online)  Amplitudes of each generalized seniority versus the excitation energy of the $K = 6$ eigenstates by calculation 2. }
\end{figure}

\begin{figure}
\includegraphics[width = 0.45\textwidth]{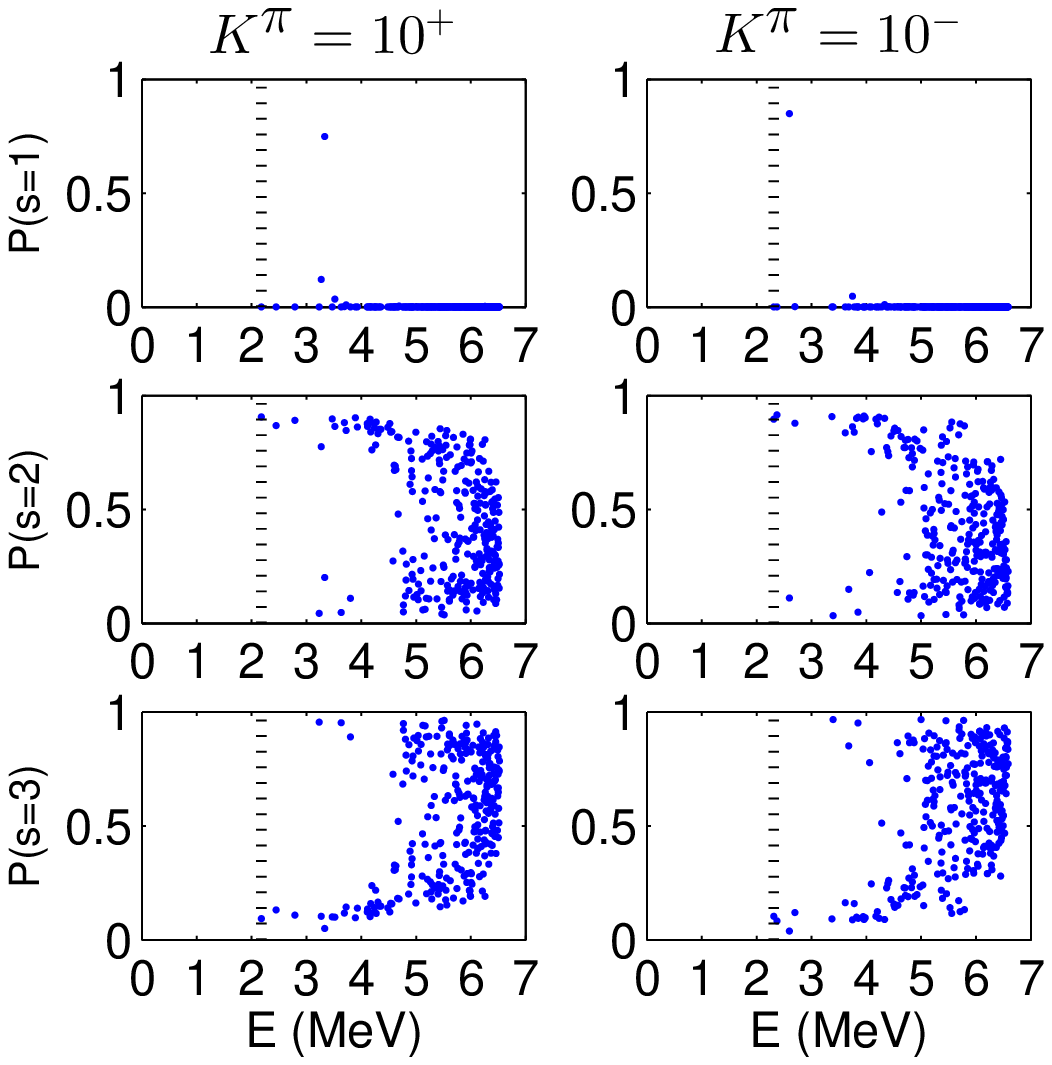}
\caption{\label{Fig_N11_s3_K10_Ps} (Color online)  Amplitudes of each generalized seniority versus the excitation energy of the $K = 10$ eigenstates by calculation 2. }
\end{figure}

The tiny $P(s=4)$ amplitudes in Figs. \ref{Fig_N9_s4_K0_Ps}-\ref{Fig_N9_s4_K10_Ps} suggest that truncating up to $S = 2s = 6$ is good enough. We do this in calculation 2 using the (larger) dimension-$46$ valence single-particle space. The results are shown in Figs. \ref{Fig_N11_s3_K0_Ps}-\ref{Fig_N11_s3_K10_Ps}. These figures are similar to Figs. \ref{Fig_N9_s4_K0_Ps}-\ref{Fig_N9_s4_K10_Ps} of calculation 1, and similar comments apply. In calculation 2 the dimension of the $|S=6, K^\pi\}$ subspace is approximately $306$, $291$, $273$, $191$, $80$ thousand for $K = 0, 2, 3, 6, 10$ and positive parity. The dimensions for negative parity are approximately the same.

It is straightforward to include proton-neutron mixing into the formalism (for example see Ref. \cite{Jia_2015}). The proton-neutron interaction, responsible for static deformations away from magic numbers, does not destroy the generalized-seniority truncation in the intrinsic body-fixed frame. Majority of the proton-neutron interaction is included in the deformed self-consistent mean field. The attractive short-range pairing force prefers coherent pairs (pairing gap around $1.5$ MeV) formed on the deformed single-particle levels. The residual interaction is not particularly strong to destroy many pairs. Work with active protons and neutrons is in progress.

%
%

\section{Conclusion}

This work proposes a fast computing scheme for generalized seniority on deformed single-particle basis. The spherical version of the algorithm \cite{Jia_2015} precalculates and stores the MPDM. Without rotational symmetry, the number of different MPDM is usually too large to fit into computer memory, and further simplification is necessary. This work analytically expresses MPDM by the normalization of the pair condensate. Precalculating and storing the normalizations instead of the MPDM greatly reduces the memory cost, and revives the algorithm. The generalized-seniority truncation of the deformed shell model runs as fast as that of the spherical shell model \cite{Jia_2015} with the new computing scheme.

The generalized-seniority truncation converges to the exact shell model when all the pairs are broken. The truncation is effective if it converges fast when only a few pairs are broken. We study the effectiveness in truncating the deformed shell model under realistic interactions by the rare-earth nucleus $^{158}_{~64}$Gd$_{94}$. The intrinsic neutron excitations (the lowest $300$ states of several multipolarity) are computed under the low-momentum {\emph{NN}} interaction $V_{{\rm{low}}-k}$, allowing as many as four broken pairs. The eigen wavefunctions are investigated in terms of amplitudes of different generalized seniority $S$. The tiny amplitudes of $S=8$ (four broken pairs) indicate the wavefunctions indeed have converged, and the truncation is very effective. Schematic pairing models usually imply that the ground state is a coherent pair condensate, and the pairs gradually break up as excitation energy increases. Our results show how well this picture survives the full realistic interaction in the intrinsic body-fixed frame. With disturbing of other parts from the realistic interaction, the pairing part remains important in forming the low-lying spectrum, and the picture remains approximately valid.


It is interesting to consider further truncation schemes on top of the generalized-seniority truncation. There are many mature truncation schemes in the shell model, such as restricting the maximal number of particle-hole excitations, cutting by mean energies of the Slater determinant basis, and more advanced techniques of selecting the basis on the fly. In fact, some of them have demonstrated effectiveness to truncate the deformed Slater determinant basis under the schematic pairing Hamiltonian \cite{Molique_1997, Zeng_1983, Zeng_1994, Xin_2000, Zhang_2011, Zhang_2012, Zhang_2013, Zhang_2013_2, Liang_2015}. These truncation schemes could be straightforwardly imposed on the unpaired nucleons of the generalized-seniority basis [the $S$ $a^\dagger$ operators of Eq. (\ref{sen_basis})], in the same way they truncate the Slater determinant basis. They further reduce the dimension, and their effectiveness with realistic interactions is interesting.

For the ground state in the intrinsic body-fixed frame, our results suggest that the variation
principle on the trial wavefunction (\ref{gs}), conventionally
called variation after particle-number projection, may
be an accurate method. Including higher-order correlations, truncation up to generalized-seniority $S = 4$ (two broken pairs) may be enough. Amplitudes of $S > 4$ are negligible. (This is consistent with the conventional wisdom that the QRPA ground state is the quasiparticle vacuum mixed mainly with the four-quasiparticle components.) It is interesting to see how general the conclusion is in other nuclei. Wherever the conclusion is valid, improving the pair condensate (\ref{gs}) by breaking two pairs can be easily done with the new algorithm in large valence spaces of over $100$ Nilsson levels. Work with active protons and neutrons is also in progress.


%
%
%


\section{Acknowledgements}

The author thanks Y. Y. Cheng and Dong-Liang Fang for discussions of using the computer code \cite{Morten_code}. Support is acknowledged from the National Natural Science
Foundation of China No. 11405109.




\newpage
~
\newpage

\end{document}